\documentclass[english,11pt,a4paper]{article}
\usepackage{babel}
\usepackage{bm}
\usepackage{makeidx}
\usepackage{layout}
\usepackage[latin1]{inputenc}
\usepackage{latexsym,amssymb,amsmath}
\usepackage{amsfonts,amsbsy}
\usepackage{eufrak}
\usepackage[mathscr]{eucal}
\usepackage{verbatim}
\usepackage{amscd}
\usepackage{xy}
\xyoption{all}
\usepackage{indentfirst}

\begin{document}
\date{}
\title{\bf {\large{ON   REGULAR   POLYTOPES  }}}
\author{Luis J. Boya\\
\small{Departamento de F\'{\i}sica Te\'{o}rica}\\
\small{Universidad de Zaragoza, E-50009 Zaragoza,
SPAIN}\\
\small{\texttt{luis.boya@gmail.com}}\\
\\
Cristian Rivera\\
\small{Departamento de F\'{\i}sica Te\'{o}rica}\\
\small{Universidad de Zaragoza, E-50009 Zaragoza,
SPAIN}\\
\small{\texttt{cristian\_elfisico@hotmail.com}}\\}\maketitle

MSC 05B45, 11R52, 51M20, 52B11, 52B15, 57S25\\

{\bf{Keywords}}: Polytopes, Higher Dimensions, Orthogonal groups

\begin{abstract}

Regular polytopes, the generalization of the five Platonic solids in
3 space dimensions, exist in arbitrary dimension $n\geq-1$; now in
{\rm dim}. 2, 3 and 4 there are \emph{extra} polytopes, while in
general dimensions only the hyper-tetrahedron, the hyper-cube and
its dual hyper-octahedron exist. We attribute these peculiarites and
exceptions to special properties of the orthogonal groups in these
dimensions: the $\mathrm{SO}(2)=\mathrm{U}(1)$ group being (abelian
and) \emph{divisible}, is related to the existence of
arbitrarily-sided plane regular polygons, and the \emph{splitting}
of the Lie algebra of the $\mathrm{O}(4)$ group will be seen
responsible for the Schl\"{a}fli special polytopes in 4-dim., two of
which percolate down to three. In spite of {\rm dim}. 8 being also
special (Cartan's \emph{triality}), we argue why there are no
\emph{extra} polytopes, while it has other consequences: in
particular the existence of the three \emph{division algebras} over
the reals $\mathbb{R}$: complex $\mathbb{C}$, quaternions
$\mathbb{H}$ and octonions $\mathbb{O}$ is seen also as another
feature of the special properties of corresponding orthogonal
groups, and of the spheres of dimension 0,1,3 and 7.
\\

\end{abstract}

\section{Introduction}

Regular \emph{Polytopes} are the higher dimensional generalization
of the (regular) polygons in the plane and the (five) Platonic
solids in space.  L. Schl\"{a}fli, studied them around $1850$ in
higher dimensions, $d > 3$ and the complete list of regular
polytopes is since then known: the existing \emph{regular} (in
particular,
\emph{convex}) polytopes, classified by dimension, are \cite{Cox-1}:\\

2d) In the plane $\mathbb{R}^2$, or inscribed in the circle
$\mathbb{S}^1$, there are regular polygons $N_2$ of \emph{any}
number of sides, $N\geq 3$; $T_2 = \{3\}$ will be the triangle and
$H_2= \{4\}$ the
square; $\{5\}$ the pentagon, etc.\\

3d) The five Platonic solids, known since the Greeks: the
Tetrahedron $\{3,3\}$, the Cube $\{4,3\}$  and dual Octahedron
($\{3,4\}$), Icosahedron and dual Dodecahedron ($Y_3$ and $Y^*_3$;
or $\{3,5\}$ and $\{5,3\}$); they can also be considered
as tessellating by curved surfaces the sphere $\mathbb{S}^2$.\\

4d) Besides the generalization of these five ($T_4$,$H_4$, $H^*_4$,
up to $Y_4$ and $Y^*_4$, called ``the $120$-cell''), there is
\emph{an extra one},
the ``$24$-cell'' ($\equiv$ ($H_4^*H$))=$\{3,4,3\}$.\\

n-d): Only the Tetrahedron, the Cube and the dual Octahedron
generalize, so $T_n$, $H_n$ and $H^*_n$ exist for any $2 \leq n <
\infty$ (although
$H_2 \approx H^*_2$), and they are the ones unique for $d >4$.\\

In the spirit of J. Stillwell's \cite{St-2} study of
\emph{exceptional objects} in mathematics, one should ask why do
these regular polytopes exist, in particular why the exceptions in
dimensions 2, 3
and 4. To understand this is the purpose of this paper, written mainly for physicists.\\

We shall argue that the special properties of the \emph{isometry}
(in particular rotation) groups in Euclidean geometry in dimensions
2 and 4 is the reason for the \emph{extra} polytopes: in dim. two
the rotation group $\mathrm{SO}(2)$ is (abelian and)
\emph{divisible}, (defined below), and in dimension four the
covering group Spin(4) = $\mathrm{SO}(4)^{\sim}$ is \emph{not
simple}, but only semisimple. Also, in dimension eight the Spin(8)
group enjoys also a special property, namely \emph{triality}, and we
shall argue why this does \emph{not} produce other regular
polytopes, but shows itself in other features.\\

The plan of this work is the following: In Sect. II we set the stage
and recall the generic construction of $T_n$, $H_n$  and $H^*_n$  in
arbitrary dimension; we shall consider the $n$-dim polytopes as
living in $\mathbb{R}^n$ and tessellating the ($n-1$)-dim sphere, as
$\mathbb{S}^{n-1}\subset \mathbb{R}^n$. We also consider the
discrete isometry and rotation invariance groups of these polytopes,
and the binary
extensions of the later groups $\mathrm{SO}(n)$ to the Spin groups, Spin($n$).\\

In Sect. III we shall consider dimension two, with the existence of
the  field of complex numbers $\mathbb{C}$ and the regular polygons
of arbitrary number of sides $N\geq 3$ a consequence of the
$\mathrm{SO}(2) =\mathrm{U}(1)\approx
\mathbb{S}^1$ group being (abelian and) \emph{divisible}.\\

In Sect. IV we shall work out the three exceptional Schl\"{a}fli
polytopes ($Y_4$, its dual $Y^*_4$ and self-dual ($H_4^*H$) ) in dim
4 as well as the quaternion division algebra $\mathbb{H}$ as a
consequence of the \emph{split} character of the Lie algebra of the
orthogonal group $\mathrm{O}(4)$; by an isotropy (stabilizer)
argument we shall also encounter the two ``old'' \emph{extra} 3-dim.
polyhedra $Y_3$ $\{3,5\}$
and $Y^*_3$ $\{5,3\}$.\\

 In Sect. V we shall (briefly) consider dimension 8,
arguing for the \emph{triality} of the Spin(8) group, with the
octonion division algebra $\mathbb{O}$, and other consequences like
parallelizability of the $\mathbb{S}^7$ sphere, and we shall also
argue why this does \emph{not} generate new polytopes.\\

Some final comments and
open questions are left for our final Sect. VI.\\

There is plenty of literature on the subject. In particular, we may
refer here to some pertinent work by Coxeter (\cite{Cox-1, Cox-3}),
\cite{Cox-4} and the recent book \cite{BB-5}). For a view of
quaternions and octonions close to ours, but not identical,
see \cite{Con-6}. The most up-to-date reference on regular polytopes is \cite{MC}.\\

\section{General Polytopes}

We take as known the precise definition of \emph{regular} polytopes
\cite{Cox-1}; we emphasize here already that we consider \emph{only
regular} polytopes, necessarily convex, and perhaps $n$-dimensional.
They can be seen existing as individual solids in $\mathbb{R}^n$,
but also more appropriately as tessellating a circumscribed sphere
$\mathbb{S}^{n-1}$ (there are, as we shall see, other spheres of
interest). The Euler characteristic $\chi$  of theses spheres is
computed (since Euler and Poincaré) as the alternative sum of the
number of faces: if the polytope has $V$ vertices, $A$ edges, $F$
surfaces, $C$ cells and in general (last) objects (boundary
polytopes) $S$, the vertices scatter through the outer sphere, the
edges become arcs of geodesics, the surfaces are curved, etc. For
this \emph{outer} sphere, containing the vertices, we have

\begin{equation}\label{eq:1}
\begin{aligned}
&\chi(\mathbb{S}^{2n})=2=V-A+F-C\textrm{\ldots}+S  \quad   \quad
\textrm{( e.g. 4 - 6 + 4 for the $T_3$)}
\\
&\chi(\mathbb{S}^{2n+1})=0=V-A+F-C\textrm{\ldots}-S  \quad
\textrm{(e.g. 5 -10 + 10 - 5 for $T_4$)}
\end{aligned}
\end{equation}

If one includes the improper sets, say the empty set and the whole
polytope, then the Euler characteristic is always 0 regardless of
whether $n$ is
odd or even.\\

Starting in dim one, the segment $I$:\quad $\!\!-\!\!\!-\!\!\!-\!$ ,
has the 0-sphere $\circ \quad \circ$ as boundary, $\partial I =
S^0=\mathbb{S}^0$: just two vertices can be considered as the only
``regular'' polytope in dimension one. One can consider also the
0-sphere as the set of norm-one real numbers in $\mathbb{R}$. Let
\emph{l} be the (arbitrary) length of
$I$, between the two circles.\\

To construct the generic regular polytopes $T_n$, $H_n$ and $H^*_n$:
we first draw (for $n=2$) in the plane $\approx\mathbb{R}^2$ the
perpendicular bisector to the segment $I$ until its distance from
the top to the two vertices equals the length \emph{l}: putting in
the edges, we have then the ordinary (plane) regular triangle
$T_2=\{3\}$. Also a parallel line to $I$ at distance \emph{l} in the
same (bi-)plane closes down to the square $H_2=\{4\}$; and if in the
triangle case, we extend symmetrically the mid-perpendicular below
$I$, we construct the rhomb as the ``2-dim'' dual of the square,
$H^*_2$, which in fact is isomorphic with the square itself
(\emph{only} in this dimension two). As $\chi(S^1)=0$, we have 3 - 3
and 4 - 4 for the vertices $V$ and edges $A$, while the rotation
symmetry groups are obviously $\mathrm{Z}_3$ and $\mathrm{Z}_4$
respectively. Denoting by $\{p\}$ the general regular polygon of $p$
edges (Schl\"{a}fli and Coxeter notations), whose existence we shall
justify, we have $p$ - $p$ as $V$ - $A$ for regular polygons, for
any $p\geq 3$. One sees also that all polygons are trivially
``self-dual'': changing vertices by edges we get the same figure;
also the (inner) circle $S^1$, tangent to the mid-edges is like the
outer one, rotated $2\pi/2N=\pi/N$ from the plane
polygon $N_2$.\\

One is easily convinced that this \emph{triple} procedure
generalizes to \emph{any} higher dimension $n > 2$, generating
hyper-tetrahedron $T_n=\{3,3,\ldots 3\}$, hyper-cube
$H_n=\{4,3,3,3\ldots 3\}$ and dual hyper-octahedron
$H^*_n=\{3,3,3\ldots 3,4\}$; for the tetrahedron $T_3$, for example,
we draw in space the normal through the centre of a regular plane
triangle $3_2 = T_2=\{3\}$ and link the top vertex with the existing
3 plane vertices to form the usual tetrahedron $T_3$; or draw a
plane parallel to the square and above it to form the cube $H_3$; or
again prolonging the normal below the plane of the triangle we get
the octahedron $H^*_3$, now $\neq H_3$. Recall trivially the number
of faces in 3 dimensions:

\begin{equation}\label{eq:2}
\begin{aligned}
&\textrm{For $T_3$:} \quad V-A+F=4-6+4; \quad \textrm{For $H_3$:}
\quad V-A+F=8-12+6; \quad\\
&\textrm{For $H^*_3$:} \quad V-A+F=6-12+8;
\end{aligned}
\end{equation}

Again, the Schl\"{a}fli-Coxeter notation  $\{p, q\}$ means (in
3-dim): surfaces of $p$ edges, and $q$ edges per vertex: hence
$\{3, 3\}$, $\{4, 3\}$ and $\{3, 4\}$ are $T_3$, $H_3$ and $H^*_3$
respectively. Here $T_3$ is self-dual, whereas $H_3$ and $H^*_3$ are
duals of each other. In these $3d$ case we have three 2-spheres, the
``middle one'' $S^2_m$ being tangent to the mid-edges.\\

The generalization to higher dimension is obvious in these three
cases. One just adds either a normal line or a parallel
(hyper-)plane to get $T_{n+1}$, $H_{n+1}$ and $H^*_{n+1}$ from the
previous, $n$-dim case. For the $T_{n+1}$, we have e.g. $n+2$
vertices connected with each other by ${n+2\choose 2}$ edges,
$\ldots $, up to ${n+2\choose n+1} = n+2$ cells $S$. Besides the
outer and the inner spheres, tangent respectively to vertices and
cells, we have several
intermediate ones (see below). For faces for $H_n$, $H^*_n$ see e.g. \cite{Cox-1}.\\

We shall \emph{not} prove (Schl\"{a}fli) that there are NO other
regular polytopes $\prod $ in arbitrary dimension, but just quote
the usual argument for three: for dim. 3 the five Platonic solids
exhaust all the possible ones (for a short argument in dim. 4 see
\cite{TER}, pag. 396): if there are $p$ edges per surface ($F$ of
them) and $q$ edges from each vertex ($V$), from Euler's $\chi= V -
A + F = 2$ we get $\frac{1}{p} + \frac{1}{q} =
\frac{1}{2}+\frac{1}{A}$, or $\frac{1}{p} + \frac{1}{q} >
\frac{1}{2}$, with the three possible solutions ($p$, $q$) = $\{3,
3\}$, $\{3, 4\}$, $\{3, 5\}$, plus the two duals, that is $T_3$,
$H_3$, $H^*_3$ and the two \emph{extra} $Y_3$, $Y^*_3$. If
$A=\infty$, $\frac{1}{p} + \frac{1}{q} = \frac{1}{2}$ would
correspond to tessellations of the (non-compact) plane
$\mathbb{R}^2$ (see Sect. VI), with solutions (Kepler): triangles
$\{3, 6\}$, dual hexagons $\{6, 3\}$ and self-dual squares $\{4, 4\}$.\\

Now for the symmetry. In Euclidean space $\mathbb{R}^n$ we can
consider the Euclidean group $\mathrm{E}(n)$, with translations,
rotations and reflections, $\mathrm{E}(n)\approx T_n \rtimes
\mathrm{O}(n)$ ($A \rtimes B$ is our conventional notation for the
semidirect product, with $A$ normal, acted upon by $B$). As
polytopes have a centre, there are no translations in their isometry
groups. Let us call $\mathrm{O}(n)$ the orthogonal group, and
$SO(n)$ the rotation subgroup: the later is a connected, compact
$\frac{n(n-1)}{2}$-dim. Lie group, $n \geq 2$. Recall, though, this
rotation group is \emph{not} simply connected, and (for $n\geq 3$)
has a double universal cover, called Spin(n). The full structure of
the orthogonal group is therefore given by

\small{\begin{equation}\label{eq:11}
\begin{CD}
@. \mathrm{Z}_2 @. @. @. @. @. @. @.
@. @.\\
@. @VVV @. @.\\
@. {\rm Spin}(n) @. @.
@. @.\\
@. @VVV @. @.\\
@. \mathrm{SO}(n) @>>> \mathrm{O}(n)
@>>> Z_2\\
\end{CD}
\end{equation}}

It is easy to see the universal cover in the 3d case (the argument
is standard in physics); namely, the ordinary rotation group
$\mathrm{SO}(3)$ contains rotations $<2\pi$ around any axis: one
parameterizes direction axis by points in $S^2$, and rotation angle
with the radius: hence $\mathrm{SO}(3)\approx B_3$, the solid 3-ball
of radius $\pi$, \emph{except} that antipodal points in the boundary
sphere are rotations differing in $2\pi$ along the same direction,
so they coincide. The topology of the rotation manifold is therefore
that of $\mathbb{R}P^3$, as the $n$-ball is like a half $n$-sphere,
and with the antipodal identification becomes the projective space.
Hence, $\pi_1(\mathrm{SO}(3))=\pi_1(\mathbb{R}P^3) = \mathrm{Z}_2$,
and one shows the result holds in general,
$n\geq 3$.\\

What about the (discrete) isometry and rotation groups for our
polytopes? It is also easy to compute them: we attack directly the
$n$-dim. case:\\

$T_n$, hypertetrahedron in $n$ dimensions: $V= n+1$, and, as any
vertex links with any other, edges  $A = {n+1\choose 2}$; surfaces
$F ={ n+1\choose 3}\ldots$; $S = {n+1\choose n}$. The Rotation group
is ${\rm Alt}_{n+1}$, the Isometry group ${\rm Sym}_{n+1}$: an
arbitrary permutation of the vertices is the most general isometry,
as any vertex connects with any other by some edge; then rotations
alone generate the even permutations, the alternative ${\rm
Alt}_{n+1}$. We can write this result in the bundle form for the
rotation part as tessellating the outer sphere, itself an
homogeneous space under the total rotation group $\mathrm{SO}(n)$
with the result

\begin{equation}
\begin{aligned}
&{\rm Alt}_n-\!\!\!-\!\!\!-\!\!\!-\!\!\!\longrightarrow {\rm Alt}_{n+1}-\!\!\!-\!\!\!\longrightarrow T_n\\
&\cap\quad \quad \quad \quad \quad \quad  \cap\quad \quad \quad \quad\cap\\
&\mathrm{SO}(n-1)\longrightarrow \mathrm{SO}(n)\longrightarrow S^{n-1}\\
\end{aligned}
\end{equation}

(The inclusions can be seen as injective arrows). Notice, for
example, that ${\rm Alt}_3 = \mathrm{Z}_3$ abelian maps injectively
into $\mathrm{SO}(2)$, abelian as well. Also, these hyper-tetrahedra
are \emph{self-dual} in the change $p$-faces by $(n-p)$ faces (see
above). In the short notation, $T_n$ is  $\{p, {p\choose 2},
\ldots,{p\choose p-1}\}$ . Notice (4) applies to vertices or the
last faces $S$, as $T_n$ is self -dual under the interchange of
$p$-faces with $(n-p)$ faces described above. But the rotation
symmetry group can act also in the intermediate faces, e.g. in the
edges: if it just permutes or flips the edges, the operation is like
a $\pi$ rotation, so the upper line above becomes
$\mathrm{Z}_2\longrightarrow {\rm Alt}_n\longrightarrow$ (edges),
and one sees the known result \cite{Cox-1} that the order of the
finite rotation group in 3d is twice
the number of edges, ${\rm |Rot\prod|}= 2A$.\\

For the hypercube $H_n$, and dual $H^*_n$ it is easy to see that the
full isometry group permutes the $n$ axes of an orthonormal frame
$\{e_1,\ldots, e_n\}$, and besides one can reflect each of them on
the origin, $e_i\longrightarrow -e_i$: for the square $\square$, the
rotation group is $Z_4$ whereas the isometry group (adding
reflections) is the dihedral $\mathrm{D}_4= \mathrm{Z}_4\rtimes
\mathrm{Z}_2$. One sees this construction permutes the $n$ basis of
a frame, plus the inversions of the same, so the isometry group is

\begin{equation}
\begin{aligned}
{\rm Isom}(H_n) = {\rm Isom}(H^*_n) = (\mathrm{Z}_2)^n\rtimes {\rm
Sym}_n \quad \quad \textrm{(order $2^n\cdot n!$)}
\end{aligned}
\end{equation}

For example, for dim 2, 3, 4 the order of the full group is 8, 48
and 384, and the order of the rotation part is half. E.g. the real
structure of the rotation groups is $Z_4$, $Sym_4$ and
$(\mathrm{Z}_2)^3\rtimes {\rm Sym}_4$, orders 4, 24 and 192 ( for
example, ${\rm Rot}(H_3)$ =${\rm Sym}_4$,
permuting the four diagonals).\\

The \emph{binary extensions} of the rotation groups were already
used back in the 19-th century \cite{Cox-3}. The Spin($n$) group is
the universal cover of $\mathrm{SO}(n)$ for $n \geq 3$. The later
notation $2\cdot G$ is used by J. H. Conway (thus the
$\mathrm{SO}(n)\cdot 2$ is the one for the full orthogonal group
$\mathrm{O}(n)$). So we can now write the binary extension in the
previous form, e.g. for the $n$-Cube $H_n$:

\begin{equation}
\begin{aligned}
&2\cdot G_0\longrightarrow \mathrm{G}_0 := \mathrm{Z}_2^{n-1}\rtimes {\rm Sym}_n -\!\!\!-\!\!\!\longrightarrow H_n\\
&\cap\quad \quad \quad \quad \quad \quad  \cap\quad \quad \quad \quad \quad \quad \quad \quad\cap\\
&{\rm Spin}(n)-\!\!\!-\!\!\!\longrightarrow \mathrm{SO}(n)-\!\!\!-\!\!\!-\!\!\!-\!\!\!-\!\!\!-\!\!\!\longrightarrow S^{n-1}\\
\end{aligned}
\end{equation}

The standard name for these 2$\cdot$ extensions is ``binary'': Thus
$2\cdot {\rm Alt}_{n+1}$ is called (for $n\geq3$) the ``binary
tetrahedral'' group,
etc.\\

The faces for the hyper-cube are easily calculated, also; for
example:\\

        For $H_4$, they are (8+8, $16\times4/2$, 24, 2+6 = 16 -32 +24 -8)\\

        For $H^*_5$, they are (10, 40, 80, 80, 32)\\

\section{Dimension two: complex numbers and polygons of $n$ sides}

The isometry group of the plane $\mathbb{R}^2$ is the Euclidean
group $\mathrm{E}_2 = \mathbb{R}^2\rtimes \mathrm{O}(2)$. For a
polygon, the centre is fixed, so the symmetry group is inside the
orthogonal group $\mathrm{O}(2)$, and the connected (rotation) part
is in $\mathrm{SO}(2) = \mathrm{U}(1)$. This group is abelian (the
general rotation group $\mathrm{SO}(n)$, $n
>2$ is not, of course), and \underline{divisible} in the category of
abelian groups, that is, under any discrete cyclic subgroup
$\mathrm{Z}_n$ the quotient is still $\mathrm{U}(1)$:

\begin{equation}
\mathrm{U}(1)/\mathrm{Z}_n\approx \mathrm{U}(1)
\end{equation}

We recall: In the category $\mathcal{A}{b}$ of abelian groups
\cite{BC}, the \emph{projective} objects are the groups
$\mathrm{Z}^n$, with $n\mathrm{Z}\approx \mathrm{Z}$, whereas the
divisible ones (= \emph{injectives}) verify
$\mathrm{U}/\mathrm{Z}_n\approx \mathrm{U}$: this is also the reason
why the ``double cover'' of the plane rotation group,
$\mathrm{SO}(2)^{~}$ is isomorphic with the original,
$\mathrm{SO}(2)$ group; recall, in this case, the universal cover
has ``infinite'' sheets, as $\pi_1(S^1) = Z$, and the covering
sequence is $\mathrm{Z}\longrightarrow \mathbb{R}\longrightarrow
S^1$. As the crucial fact for our next construction (of the complex
numbers $\mathbb{C}$ as well as the regular polygons) is that the
1-sphere $S^1$ can be endowed with a group structure, we add the
following simple argument: when $\mathrm{O}(n)$ operates in
$S^{n-1}$, the action is still transitive, with stabilizer
$\mathrm{O}(n-1)$; hence, for $\mathrm{O}(2)$ acting on $S^1$, the
isotropy group is $\mathrm{O}(1) = \mathrm{Z}_2$; going to
rotations, $\mathrm{SO}(2)$ is still transitive on the circle, with
stabilizer $\mathrm{SO}(1) = 1$. Hence $S^1\approx \mathrm{SO}(2)$,
acquires an (abelian and divisible,
as $\mathrm{SO}(2)\approx \mathrm{U}(1)$ ) group structure!\\

This produces several consequences: first, one can see the
\underline{complex numbers} as an emerging structure in
$\mathbb{R}^2$ in this way: write $\mathbb{R}^2\backslash\{0\}$ as
$S^1\times R^+$, as the polar coordinates in the plane, where
$S^1\approx \mathrm{U}(1) = \mathrm{SO}(2)$; notice the two groups
$\mathrm{U}(1)$ and $R^+$ are abelian and commute, and establish
then the multiplication law ( $\approx$ for complex numbers
$\neq$0): ($\phi$ , r)$\cdot$($\phi$´, r´) := ($\phi$ +$\phi$´ mod
2$\pi$, rr´). Add now the origin 0 with 0$\cdot$($\phi$ , r) :=0: we
have the complex numbers $\mathbb{C}$ with the usual invertible
multiplication: $(\phi,r)^{-1}=(-\phi,\frac{1}{r})$ (the addition is
the normal one for ``vectors'' in $\mathbb{R}^n$ for $n=2$). We
leave to the reader to check the
distributive law: z(z´+ z´´) = zz´+ zz´´. The moral:\\

\emph{Both} the existence of the field of complex numbers
$\mathbb{C}$ \emph{and} the existence in the plane of regular
polygons $N_2$ of \emph{any} arbitrary number of sides $N$ come from
the following fact: that the rotation group $\mathrm{SO}(2)$ is
abelian \underline{and} divisible$\ldots$ and that the 1-sphere
$S^1$ has a group structure: The abelian character of
$\mathrm{SO}(2)$ guarantees the existence of the field of complex
numbers (recall: commutativity is a property of the modern
definition of a field). But also implies the \emph{existence of the
regular polygons}: the previous equation (7) just means one can
inscribe a regular polygon of $n$ sides in the circle. This result
is so trivial that usually one never bothers asks
why regular polygons do exist with any number of sides.\\

For completeness, we express the rotation subgroup of the symmetry
group for an arbitrary (plane) polygon $N_p$, as embedded in the
rotation $\&$ orthogonal groups as
\begin{equation}
\begin{aligned}
&\mathrm{Z}_p\approx N_p\\
&\cap\quad \quad\cap\\
 &\mathrm{SO}(2)\approx S^1
\end{aligned}
\quad \quad \textrm{and}\quad \quad
\begin{aligned}
&\mathrm{Z}_2\longrightarrow \mathrm{D}_p\longrightarrow \mathrm{N}_p\\
&\|\quad \quad\quad \cap\quad \quad \quad \cap\\
 &Z_2\longrightarrow \mathrm{O}(2)\longrightarrow S^1
\end{aligned}
\end{equation}
where $\mathrm{D}_n = \mathrm{Z}_n\rtimes \mathrm{Z}_2$ is the dihedral group.\\

Although the binary groups exist, they are not the universal cover,
because
$\pi_1(\mathrm{SO}(2)) = \pi_1(S^1) = \mathrm{Z}$, and we shall not consider them.\\

\section{Dimension four: the Lie algebra of $\mathrm{O}(4)$ group splits}

For any Lie group $\mathrm{G}$ we know its Lie algebra, written
$L(G)$, which is the algebraic structure of vectors in the tangent
space to the identity of the group (a Lie group is by definition
also a manifold) propagated along the whole group by the very same
group action: so the algebra is generated by (e.g.) the
left-invariant vector fields at the identity, and has the dimension
of the group (as a real
vector space).\\

The \emph{complex simple} Lie algebras were classified by Cartan
back in 1894. He also wrote some identities among small orders, for
example $A_1 = B_1 = C_1$, that is, in the \emph{compact group} form
$\mathrm{SU}(2) = {\rm Spin}(3) = \mathrm{SpU}(1)$; the last of the
four Cartan identities says that $D_2 = B_1 + B_1$, or
Lie($\mathrm{O}(4)$) = Lie($\mathrm{O}(3)$) + Lie($\mathrm{O}(3)$).
As this is the property we need in our construction, we supply
\emph{two} different proofs of the split character of the
$L[\mathrm{O}(4)]$: an algebraic proof, and a topological one;
(Cartan's simpler proof was from the Diagram $D_2\approx \circ \quad
\circ$ (for Dynkin diagrams see e.g. \cite{Jac-9}).\\

1) \emph{Algebraic} proof: The rotation group is unimodular,
$\mathrm{SO}(n) \subset \mathrm{SL}_n(R)$; hence its representations
in $p$-forms and in ($n-p$)-forms are equivalent. On the other hand,
the Lie algebra of $\mathrm{O}(n)$ is realized by all antisymmetric
matrices, as $^too = 1\Longrightarrow L = -^tL$, where $o(t) := {\rm
exp}(tL)$. Therefore, for dim 4 and \emph{only for that}, the Lie
algebra of the group splits: Lie[$\mathrm{O}(4)$] = self-dual part
plus anti-self-dual, or 6 = 3+3, or

\begin{equation}
{\rm Lie}[\mathrm{O}(4)]\approx {\rm Lie}[\mathrm{O}(3)]\oplus {\rm
Lie}[\mathrm{O}(3)]
\end{equation}

At the level of groups, the equations are, for example

\begin{equation}
{\rm Spin}(4) ={\rm Spin}(3)\times {\rm Spin}(3)
\end{equation}

2) The \emph{topological} second proof is given in the bundle
language: write the \emph{principal bundle} $(G, B, M)$ as

\begin{equation}
\mathrm{SO}(3)\longrightarrow \mathrm{SO}(4)\longrightarrow S^3
\end{equation}
i.e., the action of $\mathrm{SO}(n)$ in the $(n-1)$ sphere for
$n=4$: the case for $n=3$ is well-known. Now the
$\mathrm{G}$-bundles $B$ over spheres $S$ are classified \cite{M-10}
by some homotopy group: if we write

\begin{equation}
G\longrightarrow B\longrightarrow S^n
\end{equation}
then the set of $G$-bundles over $S^n$ is  $H^1( S^n, G)$ in
\v{C}ech cohomology, and is given by

\begin{equation}
H^1(S^n, G) = \pi_{n-1}(G)
\end{equation}
(essentially because one trivializes the sphere by two charts
overlapping in the equator). Now a fundamental result proved by
Cartan \cite{Car-11} is that

\begin{equation}
\pi_2(G) = 0
\end{equation}
for any (finite dimensional) Lie group $\mathrm{G}$ (intuitively,
the topology of Lie groups is given by that of products of
\emph{odd} dimensional spheres \cite{Boy-12}). Hence, in our case
$\mathrm{SO}(4)\approx \mathrm{SO}(3)\times S^3$   (\emph{not} as
product of groups: $\mathrm{SO}(3)$ is not normal in
$\mathrm{SO}(4)$!) which clearly amounts to our result

\begin{equation}
{\rm Lie}[\mathrm{O}(4)] = {\rm Lie}[\mathrm{O}(3)]^2
\end{equation}
 As expected from the $2d$ case, \emph{both} the existence of the
quaternion division algebra $\mathbb{H}$ (\emph{not} a field, but a
skew-field: quaternion product (Hamilton, 1842) is not commutative!)
and the \emph{extra} Schl\"{a}fli polytopes (in 4-dim.) are a
consequence of this: Let us first show the \underline{quaternion}
division algebra; it should be already clear to the reader that

\begin{equation}
S^3\approx \mathrm{SU}(2) = {\rm Spin}(3):=
\mathrm{SO}(3)^\thicksim= 2\cdot \mathrm{SO}(3) = \mathrm{SpU}(1)
\end{equation}
(One writes $\mathrm{SU}(2)$ being unitary and unimodular, with
centre $Z_2$). So we can now sort of repeat the construction of the
complex case: we proved the crucial fact, that the 3-sphere $S^3$
admits also (like $S^1$) a (non-commutative!) group structure; the
simplest proof: $\mathrm{SU}(2)$ acts transitive in the sphere
$S^3\subset\mathbb{R}^4=\mathbb{C}^2$ with stabilizer
$\mathrm{SU}(1)={\rm I}$; next, in $\mathbb{R}^4\backslash\{0\}$ we
form the direct product of groups, as before

\begin{equation}
\mathbb{R}^4\backslash\{0\}\approx S^3\times \mathbb{R}^+
\end{equation}
with again the composition law $(\bm\theta , r) (\bm\theta', r') =(
\bm\theta\cdot \bm\theta', rr')$, and adding again the 0 as above
(in the complex case) we have formed the division algebra of
Hamilton quaternions! (In particular, the inverse of $(\bm\theta ,
r)\neq 0$ is $(\bm\theta^{-1}, \frac{1}{r})$. Here $\bm \theta$ is a
label of the 3-parameter element of $\mathrm{SU}(2)$. Hence,
$\bm\theta\cdot\bm\theta'$ is explicit in the multiplication in
$\mathrm{SU}(2)$. Of course, $S^3$ becomes then the group of
\emph{unit} quaternions, and the ``equator'' $S^2$ (no longer a
group!) can be identified with the set of unit \emph{imaginary}
quaternions also. So we see, as announced, that the quaternions owe
its existence to the Spin(4) group being only semisiple, not simple,
what makes also the 3-sphere a Lie group: Another expression of the
same thing, would be that it is the product (group) structure in the
odd spheres $S^1$ and $S^3$ which is ultimately responsible for the
field of the complex numbers ($S^1$) and the skew-field of the
quaternions ($S^3$): later we shall show how the ``sphere'' $S^0$
with just two points and the seven-sphere $S^7$ are also associated
with the real field $\mathbb{R}$ and the
octonion division algebra $\mathbb{O}$, respectively.\\

Now for the extra polytopes: In the hypertetrahedron
$T_4=\{3,3,3\}$, which exists in arbitrary dimension, we have the
rotation diagram

\begin{equation}
\begin{aligned}
&{\rm Alt}_4-\!\!\!-\!\!\!-\!\!\!\longrightarrow {\rm Alt}_5\circ\!\!-\!\!\!-\!\!\!\longrightarrow T_4\\
&\cap\quad \quad \quad \quad \quad \cap\quad \quad \quad \quad \cap\\
&\mathrm{SO}(3)-\!\!\!\longrightarrow \mathrm{SO}(4)-\!\!\!\longrightarrow S^3\\
\end{aligned}
\end{equation}
As $\mathrm{SO}(4)$ splits, the image of ${\rm Alt}_5$, which is a
\emph{simple} group, must go in one of the factors, so it must be
$\mathrm{SO}(3)$, which is (sub)group: \emph{we have then a regular
polytope in 3-dim with rotation symmetry} ${\rm Alt}_5$, so we have
a new polytope in 3 dimensions! (notice $Z_4$ is not contained in
${\rm Alt}_5$):

\begin{equation}
\begin{aligned}
&\mathrm{Z}_i-\!\!\!-\!\!\!-\!\!\!\longrightarrow {\rm Alt}_5\circ\!\!-\!\!\!-\!\!\!\longrightarrow Y_3\\
&\cap\quad \quad \quad \quad \quad \cap\quad \quad \quad\cap\\
&\mathrm{SO}(2)-\!\!\!\longrightarrow \mathrm{SO}(3)-\!\!\!\longrightarrow S^2\\
\end{aligned}
\end{equation}
($\mathrm{Z}_i$, $i=2,3$ or 5) That is to say, the action
$\mathrm{SO}(3)$ to $S^2$ has the abelian $\mathrm{SO}(2)$ as
stabilizer, hence one should ask for an abelian subgroup of ${\rm
Alt}_5$, and the two more useful ones are $\mathrm{Z}_3$ and
$\mathrm{Z}_5$: this is the rotation symmetry of either triangular
or pentagonal faces, so we have constructed both the icosahedron
$Y_3$ and the dual dodecahedron $Y^*_3$ ! Notice that we have to use
the fact that polygons of arbitrary (five, three $\ldots$) number of
edges exist in dimension two! Also, the subgroup $\mathrm{Z}_2$
corresponds to the involutions on the edges, see just below.\\

We generate therefore the icosahedron $Y_3=\{3,5\}$, with $V - A + F
= 12 - 30 + 20$, and the dual dodecahedron $Y^*_3=\{5,3\}$ with $20
- 30 + 12$. Notice, as $V + F = 32$ in both cases, the number of
edges is fixed to 30: the two exotic Platonic solids are being
explained (we shall see them again as coming from four dimensions!).
Notice even the subgroup $\mathrm{Z}_2$ in ${\rm Alt}_5$: it occurs
when the ${\rm Alt}_5$ group acts on
the 30 ($=\frac{60}{2}$) edges of either $Y_3$ or $Y^*_3$.\\

But there is more: the extra polytopes in dimension four!\\

As $Alt_5$ lies in $\mathrm{SO}(3)$, it can be ``lifted'' to $2\cdot
Alt_5$, order 120, by going to the covering $S_3\approx
\mathrm{SU}(2)$ = Spin(3). Hence we have a 120-vertex tessellation
of $S^3$, one of the Schl\"{a}fli new polytopes! The situation is
this:

\begin{equation}
\begin{aligned}
&{\rm Alt}_5-\!\!\!-\!\!\!-\!\!\!\longrightarrow G_{120\times60}-\!\!\!-\!\!\!\longrightarrow Y_{120}\\
&\cap\quad \quad \quad \quad \quad \cap\quad \quad \quad \quad \cap\\
&\mathrm{SO}(3)-\!\!\!\longrightarrow \mathrm{SO}(4) -\!\!\!\longrightarrow S^3\\
\end{aligned}
\end{equation}

But ${\rm Alt}_4$ acts on the $T_3$ polytope in 3d, so it is also a
subgroup of $\mathrm{SO}(3)$. Then

\begin{equation}
\begin{aligned}
&{\rm Alt}_4-\!\!\!-\!\!\!-\!\!\!\longrightarrow G_{600\times24}-\!\!\!-\!\!\!\longrightarrow Y_{600}\\
&\cap\quad \quad \quad \quad \quad \cap\quad \quad \quad \quad \cap\\
&\mathrm{SO}(3)-\!\!\!\longrightarrow \mathrm{SO}(4)-\!\!\!\longrightarrow S^3\\
\end{aligned}
\end{equation}
The construction is similar to the icosahedron/dodecahedron in
$3d$:\\

To complete the enumeration of the faces, we argue in the following
way: from the $Y_3$, the $3d$ icosahedron, $12 - 30 + 20$, we have
$20\times \frac{3}{2}= 30$ edges, so now the number of edges per
vertex is one more, and we have $600\times \frac{4}{2}= 1200$ edges,
so, as $\chi =0$, the two new potytopes have:
\begin{equation}
\begin{aligned}
Y_4: 120, 720, 1200, 600,\quad \textrm{and}\quad Y^*_4\quad
\textrm{the dual},\quad  600, 1200, 720, 120.
\end{aligned}
\end{equation}

In other words, we ``understand'' both the 120-cell and the 600
cell. We can even retrieve the $3d$ polyhedra $Y_3$ and $Y^*_3$ by
fixing a vertex, getting $12 -30 +20$ for $Y_3$ and $20 -30 +12$ for
the dual $Y^*_3$, as we know already. The real novelty is the
appearance of the pentagons, both in $3d$ and in $4d$. In
Schl\"{a}fli notation, we have\\

$\{3, 3, 3\}$ for $T_4$, $\{4, 3, 3\}$ and dual $\{3, 3, 4\}$ for
$H_4$ and $H^*_4$.; $\{3, 3, 5\}$ for the 600 cell $Y_4$, and dual
$\{5, 3, 3\}$ for
the 120-cell, $Y^*_4$.\\

So we ``understand'' five of the six regular polytopes in 4
dimensions. What about the remaining one, called ``the 24-cell''?
Here, unfortunately, we do not have such a ``pentagonal'' argument,
but a simpler one, characteristic also of dimension four, which does
NOT descend to three. This is our construction:\\

Compare the hypercube $H_4$ ( +16 -32 +24 -8) with the dual
hyper-octahedron $H^*_4$ (8 -24 + 32 -16): the relation, in the
square, between centre and half-side vs. centre and corner is
$\sqrt{2}$. In the ordinary cube, is $\sqrt{3}$ and in the $H_4$ is
2 $= \sqrt{1+1+1+1}$: Hence, the vertices of the hyperoctahedron are
at the same distance (2) from the vertices of the hypercube
themselves!. Hence, a \emph{unique} new polytope is possible in four
dimensions, made out of the two: each vertex combines with four
others form the $h$-cube, and with another 4 from the
$h$-octahedron, to produce 8+16 vertices, and $24\cdot\frac{8}{2}$
edges, etc.:

\begin{equation}
\begin{aligned}
\textrm{The 24-cell}:\quad (24, 96, 96, 24),\quad \textrm{which is self-dual}\\
\end{aligned}
\end{equation}

In the Coxeter-Schl\"{a}fli notation, the 24-cell is clearly $\{3,
4, 3\}$. It only remains to compute the isometry groups: one sees
``easily'' that one should include three new rotations, for the
triangular surfaces of
the octahedron:\\

$|{\rm Isom} (24-cell) | = 3\cdot |{\rm Isom} (H_4)| = 1152$\\
which coincides, as it should, with the Weyl Group of the second
exceptional Lie Group $F_4$ (private communication from Joseph C.
Varily).

\section{Dimension 8: triality}

Dimension 8 is also special (if only because the existence of
octonions!). The characteristic property is \emph{triality}, already
discovered by \'{E}. Cartan in 1925; namely, the centre of the
$\mathrm{SO}(8)$ group is $\mathrm{Z}_2$, as in any even dimensional
rotation group; so the Spin(8) group has centre of order four, again
as in any Spin(2n) group: but, from Bott periodicity \cite{MM}, we
know: for Spin(4n+2) the centre is $\mathrm{Z}_4$ (e.g. Spin(6) =
$\mathrm{SU}(4)$), and for Spin(4n) the centre is $\mathrm{V} =
(\mathrm{Z}_2)^2$. For dim $8n+4$ the two \emph{spinor}
representations are \emph{quaternionic}, that is, complex, and
conjugate equivalent, but not possible in real form. But in dim $8n$
they are both real, and for dim 8 (and only for that!) the dim is
also $8= 2^{\frac{8}{2}-1}$: the three \emph{irreps} namely
vectorial $\square$ , and the two chiral ones, $\vartriangle_L$ and
$\vartriangle_R$ are real and of dimension 8: they can be
interchanged under a $S_3$ symmetry: that is the original Cartan's
triality (on the other hand, a very obvious symmetry from the Dynkin
diagram). Notice the quoted centre $V=(Z_2)^2$ also plays a role:
these three irreducible representatations are not faithful, with a
$\mathrm{Z}_2$ kernel, different in each case: the three
$\mathrm{Z}_2$ subgroups of the center group, $\mathrm{V} =
(\mathrm{Z}_2)^2$ (in any group $\mathrm{G}$, if there are more than
one central involution, no \emph{irrep} can be faithful, as Schur's
lemma forces these involutions to be represented by $\pm 1$). It is
remarkable that $Aut(Center = \mathrm{V}) = S_3$
=\emph{Out}(Spin(8)), but the two groups seem to have nothing to do
with each other! This, in spite of the morphism between
\emph{outo}morphisms of the group and automorphisms of the center:
Out(G)$\longrightarrow
{\rm Aut}(\mathrm{Z}_G)$ for \emph{any} group $\mathrm{G}$.\\

There is a slight generalization of triality, discussed by J. F.
Adams \cite{Ad-13}; namely in any algebra, there is a product in the
underlying vector space: $V\times V\longrightarrow  V$. In the form
$V\times   V\times   V\longrightarrow 1$, it is Adams's triality: in
particular it holds for the four division algebras $\mathbb{R}$,
$\mathbb{C}$, $\mathbb{H}$ and $\mathbb{O}$. In our case, one shows
easily that $\square \times \vartriangle_L \times \vartriangle_R$
contains the identical representation just once (octonion
multiplication).\\

As the reader shall recall, the extra polytopes in dim 4 owe their
existence to the fact that the Spin(4) is a direct product, hence
the Rot groups are also a direct product of spaces: none of this
happens in Spin(8), so NO new regular polytopes arise. Nevertheless,
there are important quasi-regular lattices in $\mathbb{R}^8$!. See
e.g. \cite{Con-16}.\\

However, the octonions do arise! How? Precisely because of  the
triality in the the sense of Adams! As octonions are not our major
object of study, we just sum up some important properties: we repeat
the conventional construction (Hamilton, Cayley, Graves): namely,
take three independent anti-involutive units $e_1$, $e_2$ and $e_3$,
so $e{_1}^2=-1$ etc; and \emph{force} all products to be still
anti-involutive, namely $e_4:= e_1e_2$, $e_5 = e_2e_3$ and $e_6 =
e_3e_1$: so they have to ANTIcommute. But there is an extra one,
$e_7 :=e_1(e_2 e_3)$: for this to be also antiinvolutive,
associativity is lost: $e_1(e_2e_3) = - (e_1e_2)e_3$ ! And we have a
base for the imaginary octonions: $e_1,\ldots, e_7$. With unit 1 =
$e_0$ and all \emph{real} combinations, we form an algebra which is
8-dim over the reals $\mathbb{R}$; a generic element would be $o
=\sum _{i=0}^7 x_i e_i$; define the conjugate $\bar{o}$ as in
$\mathbb{C}$ or $\mathbb{H}$: changing the signs but of first term.
Then for $o\neq0$ the norm $\bar{o}o = \mathcal{N}(o)$ is real $\neq
0$, so define the inverse as $o^{-1} = o /\mathcal{N}(o)$, and one
has the (nonassociative, noncommutative) division algebra of the
octonions!\\

In particular, norm-one octonions form a multiplicative structure
(with inverse) in $S^7$, but \emph{not} a group because of the lack
of associativity. And it is satisfactory that, back in 1958, Bott
and Milnor proved \cite{Mil-14}, by topological arguments, that
$S^0(=Z_2 = O(1))$, $S^1 (\approx \mathrm{U}(1)=\mathrm{SO}(2))$,
$S^3 ( \approx Sq(1)=SU(2)=Spin(3))$ and $S^7$ were the only
\emph{parallelizable} spheres, clearly associated, respectively, to
$\mathbb{R}$, $\mathbb{C}$, $\mathbb{H}$ and $\mathbb{O}$.

\section{Final comments}

There are several related things we left behind. For example,
tessellations in other spaces: spheres $S^n$ are characterized by
being of constant \emph{positive} curvature and hence they are
compact. What about other constant curvature spaces, in concrete
\emph{flat} (i.e. $\mathbb{R}^n$) and \emph{negative curvature}
(e.g. the hyperbolic space $H^n$)? In the mathematical literature
they are called ``maximally symmetric spaces'' \cite{Hel-15}. We
just recall a couple of well-known results:

\begin{itemize}
  \item a)  In $\mathbb{R}^2$, we already realized the regular tessellation condition $\frac{1}{p} + \frac{1}{q} =\frac{1}{2}$,
            with the three solutions by triangles, squares and hexagons: it is easy to connect the three with the special character
            of the rotation isometry group, $\mathrm{E}^+(2) = \mathbb{R}^2\rtimes \mathrm{SO}(2)$, which is nilpotent (semidirect product of two
            abelian groups): In any other dimension $\neq 4$, only (hyper-)cubes $H_n$ tessellate the Euclidean space
            (the 24-cell and the dual $H^*_4$ tessellate $\mathbb{R}^4$ also).

  \item b)  For $H^2$, hyperbolic, the restriction condition is $\frac{1}{p} + \frac{1}{q} <\frac{1}{2}$, with infinite solutions, like e.g.
 $ \{7, 3\}$ : heptagons, three per vertex; as the space is non-compact, with arbitrary (constant negative) curvature, there are
 infinite tessellations.
\end{itemize}

Are there other manifolds which can support regular polytopes? For
example, what about symmetric rank-one spaces, like $CP^n$? At the
moment, we have nothing (new) to say about this situation, but please check \cite{MC}.\\

As a final reference, the review by Conway and Sloane \cite{Con-16}
contains much material related to the one dealt with here.

\end{document}